\documentstyle[eqsecnum,aps,preprint]{revtex}  
\begin{document}                
\title{Group Theoretical Approach to the Coherent and the Squeeze States of
a Time-Dependent Harmonic Oscillator with a Singular Term}
\author{Jung Kon Kim and Sang Pyo Kim}
\address{Department of Physics, \\
Kunsan National University, \\
Kunsan 573-360, Korea}
\preprint{KNU-TH-16 \\ April 1994}
\maketitle
\begin{abstract}
For a time-dependent harmonic oscillator with an inverse squared singular
term, we find the generalized invariant using the Lie algebra of $SU(2)$
and construct the number-type eigenstates and the coherent states using the
spectrum-generating Lie algebra of $SU(1,1)$. We obtain the evolution
operator in both of the Lie algebras. The number-type eigenstates and
the coherent states are constructed group-theoretically for both the
time-independent and the time-dependent harmonic oscillators with the singular
term. It is shown that the squeeze operator transforms unitarily the
time-dependent basis of the spectrum-generating Lie algebra of $SU(1,1)$
for the generalized invariant, and thereby evolves the initial vacuum into
a final coherent vacuum.

\end{abstract}

\pacs{03.65.Fd, 02.20.+b}

\section{Introduction}               
The adiabatic method has been one of the most general and frequently used
approximate methods for time-dependent quantum systems \cite{1}.
The task to find the exact quantum states of these systems, however, is
mathematically difficult in general. The exact quantum states are known
only for a few systems. The solutions for the physically interesting,
nonlinear,
time-dependent classical harmonic oscillators even dated back to
Ermakov \cite{2}, and they are now referred to as the Ermakov system.

There are several methods to find the exact quantum states for such
time-dependent quantum systems. The first method which was developed several
decades ago by Lewis and Riesenfeld \cite{3} introduces an interesting
quantum-mechanically conserved quantity, now known as either the
Lewis-Riesenfeld invariant or the generalized invariant,
for a time-dependent quantum harmonic
oscillator and finds the exact quantum states in terms of the eigenstates
of the invariant up to some time-dependent phase factors. The Ermakov
system also has a generalized invariant called the Lewis-Ray-Reid
invariant \cite{4}. The second method which was developed by Wei and Norman
\cite{5} finds directly the evolution operator in the disentangled exponentials
for a time-dependent quantum system that has some Lie algebraic structure.
The path integral \cite{6} is still another method.

 As one of the exactly solvable time-dependent quantum systems, harmonic
 oscillators have been studied intensively and have had a wide application
 for a long period of time in various branches of physics  from quantum
 optics to gravitational wave detection. Among the various
 methods introduced to find the exact quantum states
 of time-dependent harmonic oscillators, the most frequently
 used methods are the generalized invariant method
 and the evolution operator method that are based on the Lie algebras of
 $SU(2)$ and $SU(1,1)$.
  First, the generalized invariant method was originally introduced for a
  time-dependent quantum harmonic oscillator with the aid of an auxiliary
  equation \cite{3}. There are further elaborations and diverse
  applications of this method \cite{7}. The second method is the
  evolution operator method in disentangled exponentials
  \cite{5} as  further elaborated in
  Ref. \cite{8}. Using one variant of the generalized invariant method
  one of the authors (S.P.K) has recently found a connection between
  the   classical and the quantum harmonic oscillators,
  and has obtained explicitly a
  class of exactly solved time-dependent quantum harmonic oscillators
  \cite{9}. Quite independently of these two methods, Popov and Perelomov have
  found the exact quantum states of a time-dependent quantum harmonic
  oscillator in the Gaussian form using the classical integrals of motion
  \cite{10}.

In this paper, we shall extend the group-theoretical approach
to a time-dependent
harmonic oscillator with an inverse squared singular term (hereafter,
anharmonic oscillator will refer to harmonic  oscillator with
an inverse squared singular term). The time-dependent
quantum anharmonic oscillator has already been analyzed
as an uncoupled Ermakov
system \cite{4} and was independently treated in Ref.\cite{11}. In
particular, the time-independent anharmonic oscillator can be regarded as
a reduced system of the Calogero model in the center of mass system
\cite{12}. The main purpose of this paper is to uncover the underlying
group structure for the time-dependent anharmonic oscillator by introducing
the two relevant Lie algebras of $SU(2)$ and $SU(1,1)$, and to construct the
exact quantum states group-theoretically. It is shown that the Lie algebra
of $SU(2)$ is useful in finding the generalized invariant, whereas the
spectrum-generating Lie algebra of $SU(1,1)$ yields the number-type
eigenstates and the coherent states. It is observed that in the Lie
algebra of $SU(2)$ the generalized
invariants for both the time-dependent harmonic oscillator and the
time-dependent anharmonic
oscillator have the same invariant equation,
the only modification being in the representation of the Lie algebra,
whose solutions can be read  from the classical integrals of
motion for the time-dependent harmonic oscillator \cite{9}. We also obtain
the evolution operator in both of the Lie algebras, which is again
expressed by the classical integrals as expected. Because the exact
quantum states are determined by the eigenstates of the generalized
invariant up to some time-dependent phase factors,
we introduce the spectrum-generating Lie
algebra of $SU(1,1)$ for the generalized invariant and construct the
number-type eigenstates and the coherent states. Finally, it is found that
the squeeze operator is simply the Hermitian conjugate of the evolution
operator, transforms unitarily the basis of the Lie algebra of $SU(1,1)$
for the generalized invariant, and thereby evolves the initial vacuum into
a coherent vacuum.

The organization of this paper is as follows: In Sec. II, we
introduce the two Lie algebras of $SU(2)$ and $SU(1,1)$ for a
time-independent anharmonic oscillator. In Sec. III, we find the generalized
invariant and the evolution operator for a time-dependent anharmonic
oscillator. In Sec. IV, we construct the number-type eigenstates and the
coherent states for both the time-independent and the time-dependent anharmonic
oscillators. In Sec. V, we show that the squeeze operator transforms
unitarily the basis of the Lie algebra of $SU(1,1)$ and is the Hermitian
conjugate of the evolution operator.

\section{$SU(2)$ AND $SU(1,1)$ GROUPS OF THE ANHARMONIC OSCILLATOR}

 A time-independent quantum anharmonic oscillator,
 \begin{equation}
 \hat{H} = \frac{\hat{p}^2}{2} + \frac{\omega^2_0 \hat{q}^2}{2} +
 \frac{c}{\hat{q}^2}      \label{2.1}
 \end{equation}
 where $c$ is a constant,  has the Lie algebra of $SU(2)$,
 \begin{equation}
 \left[ \frac{i}{2} \hat{L}_0,\hat{L}_{\pm} \right] = \pm \hat{L}_{\pm} ,
 \left[ \hat{L}_+ , \hat{L}_- \right] = 2 \left( \frac{i}{2} \hat{L}_0
 \right) , \label{2.2}
 \end{equation}
with the following choice of the Hermitian basis:
\begin{equation}
\hat{L}_- = \frac{\hat{p}^2}{2} + \frac{c}{\hat{q}^2} , \hat{L}_0 =
\frac{\hat{p} \hat{q} + \hat{q} \hat{p}}{2} , \hat{L}_+ =
\frac{\hat{q}^2}{2} . \label{2.3}
\end{equation}
One may introduce the standard basis for the Lie algebra of $SU(1,1)$ as
\begin{equation}
\hat{K}_0 = \frac{1}{2} \left( \hat{L}_- +\hat{L}_+ \right) ,
\hat{K}_{\pm} = \frac{1}{2} \left( \hat{L}_+ - \hat{L}_- \mp i \hat{L}_0
\right) ,                        \label{2.4}
\end{equation}
whose group structure is
\begin{equation}
\left[ \hat{K}_0, \hat{K}_{\pm} \right] = \pm \hat{K}_{\pm} , \left[
\hat{K}_+, \hat{K}_- \right] = - 2 \hat{K}_0 . \label{2.5}
\end{equation}
In particular, one can represent the standard basis of the $SU(1,1)$
algebra as
\begin{equation}
\hat{K}_0 = \frac{1}{2} \left( \hat{a}^\dagger \hat{a} + \frac{1}{2} \right) +
\frac{c}{2 \hat{q}^2}, \hat{K}_+ = \frac{1}{2} \hat{a}^{\dagger 2}
- \frac{c}{2 \hat{q}^2} , \hat{K}_- = \frac{1}{2} \hat{a}^2
- \frac{c}{2 \hat{q}^2}, \label{2.6}
\end{equation}
in terms of the creation and the annihilation operators of a harmonic
oscillator. It should be noted that the anharmonic oscillator
is one of the two perturbations of the harmonic
oscillator which have finite-dimensional Lie algebras (the other being that
with a linear force term). All other extended
harmonic oscillators with perturbation
terms except for those with
the linear force term and the inverse squared
singular term have
infinite-dimensional Lie algebra, for which the Weyl ordered basis in  Ref.
\cite{13} may be used. The bases in Eqs. (\ref{2.3}) and (\ref{2.4}) are chosen
from among the bases of such Lie algebras for a later use.
 One can also use the following basis:
\begin{equation}
\hat{\Gamma}_0 = \frac{1}{2} \hat{L}_0 , \hat{\Gamma}_\pm = -2 \hat{L}_-
\pm \frac{1}{8} \hat{L}_+ ,         \label{2.7}
\end{equation}
as a spectrum-generating algebra of $SU(1,1)$ \cite{14}.

\section{ TIME-DEPENDENT QUANTUM ANHARMONIC OSCILLATOR}

We now turn to a  time-dependent quantum anharmonic oscillator
of the form of Eq. (\ref{2.1}), but with a variable frequency squared:
\begin{equation}
\hat{H} = \frac{\hat{p}^2}{2} + \frac{\omega^2 (t) \hat{q}^2}{2} +
\frac{c}{\hat{q}^2}. \label{3.1}
\end{equation}
The classical equation of motion is the equation for the anharmonic
oscillator:
\begin{equation}
\ddot{q} + \omega^2 (t) q - \frac{2 c}{q^3} = 0.  \label{3.2}
\end{equation}

The time-dependent anharmonic oscillator was first analyzed by Camiz {\it et
al}. \cite{11} and independently in Ref. \cite{6}. The Lewis-Ray-Reid invariant
was used for the time-dependent anharmonic oscillator as an uncoupled
Ermakov system \cite{4}. The exact eigenfunctions in Gaussian form up to some
time-dependent factors are similar in many respects to those for the
Caldirola-Kanai
oscillator \cite{15}.
In this paper we shall, however, uncover the underlying group structure
introduced in Sec.~II and elaborate further on the coherent and squeezed
 states of the time-dependent anharmonic oscillator. Two methods, the
generalized invariant method and the evolution operator method,
will be employed below for the time-dependent anharmonic oscillator
 of Eq. (\ref{3.1}).

\subsection{Generalized Invariant Method}

The generalized invariant method  finds quantum-mechanical invariants
for a time-dependent quantum system and  obtains the exact quantum states
as the eigenstates of the invariant. In the Heisenberg picture, the
quantum-mechanical invariants obey the equation (in units of $\hbar=1$ )
\begin{equation}
\frac{d}{dt} \hat{I} (t) = \frac{\partial}{\partial t} \hat{I} (t) - i
\left[ \hat{I} (t) , \hat{H} (t) \right] = 0 .  \label{3.3}
\end{equation}
Since the invariant equation is linear, the quantum-mechanical invariants
form a linear space of operators. In particular, when the Hamiltonian has Lie
algebras, the quantum-mechanical invariants belong to the same Lie
algebras. In the case of a time-dependent harmonic oscillator, the
Hamiltonian has the Lie algebras of $SU(2)$ and $SU(1,1)$, therefore
the generalized
invariant is a quantum-mechanical invariant that has the same Lie
algebras of $SU(2)$ and $SU(1,1)$. The generalized invariant is determined
uniquely up to a one-parameter coefficient, so one
gets the same generalized invariant, regardless of the basis used, as
long as the basis spans the same Lie algebra as the Hamiltonian. In our case of
the
anharmonic oscillator, we shall get the same generalized invariant
irrespective of whether the basis in Eq. (\ref{2.3}) or
the basis in Eq. (\ref{2.4}) is used.

We use the basis in Eq. (\ref{2.3}) rather than the basis in Eq. (\ref{2.4})
in analogy to the harmonic oscillator case, and we find a generalized invariant
of the  form
\begin{equation}
\hat{I} (t) = \sum_{\scriptstyle k=0,\pm \atop} g_k (t) \hat{L}_k.
\label{3.4}
\end{equation}
The invariant equation can be written as a vector equation:
\begin{equation}
\frac{d}{d t} \left(\begin{array}{clcr}
                           g_- (t)  \\
                           g_0 (t)  \\
                           g_+ (t)  \\
                    \end{array} \right)
               = \left( \begin{array}{clcr}
                 0            &  -2            &     0   \\
                 \omega^2 (t) &  0             &     -1   \\
                 0            &2 \omega^2 (t)  &      0   \\
                \end{array} \right) \left(\begin{array}{clcr}
                                                 g_- (t)  \\
                                                 g_0 (t)  \\
                                                 g_+ (t)  \\
                                          \end{array} \right) .
\label{3.5}
\end{equation}
It is worthy  to compare Eq. (\ref{3.5}) with Eq.
(11) of Ref. [9(a)] and Eq. (3.5) of Ref. [9(c)] for the
unperturbed time-dependent quantum harmonic oscillator
\begin{equation}
\hat{H}_0 (t) = \frac{\hat{p}^2}{2} + \frac{\omega^2 (t) \hat{q}^2}{2} .
\label{3.6}
\end{equation}
The only differences between these two cases are the elements $\hat{L}_- =
\hat{p}^2 / 2 + c / \hat{q}^2 $ of $SU(2)$ for the anharmonic oscillator
and $ \hat{L}_- = \hat{p}^2 / 2 $ (i.e, $c = 0 $) for the harmonic
oscillator. In either case, both of the Hamiltonians have the same Lie
algebraic structure
\begin{equation}
\hat{H} (t) = \hat{L}_- + \omega^2 (t) \hat{L}_+ ~~.       \label{3.7}
\end{equation}
Therefore, one may expect the same invariant equation.
However, it is only through
a group-theoretical approach that one can manifestly explain the reason
for the same invariant equation (Eq. (\ref{3.5}))
or the same auxiliary equation \cite{3}
\begin{equation}
\ddot{\rho} (t) + \omega^2 (t) \rho (t) = \frac{1}{\rho^3 (t)}
\label{3.8}
\end{equation}
by using  the following substitution as was done in Refs. [9(a)] and [9(c)]:
\begin{equation}
g_- (t) = \rho^2 (t)~~,~~g_0 (t) = - \rho (t) \dot{\rho} (t)~~,~~ g_+ (t) =
\dot{\rho}^2 (t) + \frac{1}{\rho^2 (t) }. \label{3.9}
\end{equation}
{}From the connection [9(a), (c)] between the classical and quantum harmonic
oscillators, the generalized invariant is given straightforwardly by
\begin{equation}
              \left(\begin{array}{clcr}
                           g_- (t)  \\
                           g_0 (t)  \\
                           g_+ (t)  \\
                    \end{array} \right)
               = \left( \begin{array}{clcr}
      Q^2_0            &  -2 Q_1 Q_0         &  Q^2_1     \\
      -P_1 Q_0         & P_0 Q_0 + P_1 Q_1   & -P_0 Q_1   \\
       P^2_1            &-2 P_1 P_0          & P^2_0     \\
                \end{array} \right) \left(\begin{array}{clcr}
                                                 g_- (t_0)  \\
                                                 g_0 (t_0)  \\
                                                 g_+ (t_0)  \\
                                          \end{array} \right)
                         \label{3.10}
\end{equation}
where
\begin{equation}
              \left(\begin{array}{clcr}
                           p(t)  \\
                           q(t)  \\
                      \end{array} \right)
               = \left( \begin{array}{clcr}
         P_0 (t,t_0) & P_1 (t,t_0)  \\
         Q_1 (t,t_0) & Q_0 (t,t_0)  \\
                         \end{array} \right) \left(\begin{array}{clcr}

                     p(t_0)  \\
                                               q(t_0)  \\
                                          \end{array} \right)
\label{3.11}
\end{equation}
represents the classical integrals of motion for the harmonic oscillator
in Eq.(\ref{3.6}).

\subsection{Evolution Operator Method}

The evolution operator method evaluates directly the evolution operator for
a time-dependent quantum system with a Lie algebraic structure. There are
the two typical methods to express the evolution operator. One method is
the global exponential operator of Magnus \cite{16}. The other is the
product of disentangled exponential operator introduced by Wei and
Norman \cite{5},
which has recently been applied to time-dependent harmonic
oscillators \cite{8}. We shall follow the technique developed in Ref.\cite{8}.

The evolution operator for the quantum anharmonic oscillator in Eq. (\ref{3.1})
obeys the evolution equation
\begin{equation}
i \frac{\partial}{\partial t} \hat{U} (t,t_0) = \hat{H} (t) \hat{U}
(t,t_0) \label{3.12}
\end{equation}
with the initial value $\hat{U} (t_0,t_0) =1$. The evolution operator is
unitary for a Hermitian Hamiltonian.
We search for the disentangled evolution operator of the form
\begin{equation}
\hat{U} (t,t_0 ) = \exp{\left( i l_+ (t) \hat{L}_+ \right)} \exp{ \left( l_0
(t)
\frac{i}{2} \hat{L}_0 \right)} \exp{\left( i l_- (t) \hat{L}_- \right)}
\label{3.13}
\end{equation}
in the basis of Eq. (\ref{2.3}) and of the form
\begin{equation}
\hat{U} (t,t_0 ) = \exp{ \left( i k_+ (t) \hat{K}_+ \right)} \exp{ \left(k_0
(t)
\hat{K}_0 \right)} \exp{ \left( i k_- (t) \hat{K}_- \right)  }
\label{3.14}
\end{equation}
in the basis of Eq. (\ref{2.4}). For the evolution operator
in Eq. (\ref{3.13}),
 we have a set of coupled differential equations:
\begin{eqnarray}
\dot{l}_+ (t) + l^2_+ (t) + \omega^2 (t) = 0 \nonumber \\
\dot{l}_0 (t) + 2 i l_+ (t) = 0 \nonumber \\
\dot{l}_- (t) + i \exp{l_0 (t)} = 0  \label{3.15}
\end{eqnarray}
with the initial values $ l_0 (t_0) = l_{\pm} (t_0) = 0 $.
One solution is found
to be
\begin{equation}
l_+ (t) = \frac{\dot{x}(t)}{x(t)} \label{3.16}
\end{equation}
where
\begin{equation}
\ddot{x}(t) + \omega^2 (t) x(t) = 0 . \label{3.17}
\end{equation}
$x(t)$ satisfies the classical equation of motion for the harmonic
oscillator in Eq. (\ref{3.6}), as expected \cite{8}.
The other solutions are obtained
by substituting Eq. (\ref{3.16}) into Eq. (\ref{3.15}) and
integrating.

For the evolution operator in Eq. (\ref{3.14}), we have the following set of
differential equations:
\begin{eqnarray}
\dot{k}_+ (t) + \frac{1}{2} \left( \omega^2 (t)  - 1 \right) ( k_+^2 -1 ) -
i \left( \omega^2 (t) + 1 \right) k_+ (t) = 0 , \nonumber \\
\dot{k}_0 (t) + \left( \omega^2 (t) -1 \right) k_+ (t) -i \left( \omega^2 (t)
+ 1 \right) = 0,  \nonumber \\
\dot{k}_- (t) - \frac{1}{2} \left( \omega^2 (t) - 1 \right) \exp{k_0 (t) }
= 0  \label{3.18}
\end{eqnarray}
with the initial values $k_0 (t_0 ) = k_{\pm} (t_0) = 0$. One solution is found
to be
\begin{equation}
k_+ (t) = \frac{2}{\omega^2 (t) -1} \frac{\dot{y}(t)}{y(t)}~~,  \label{3.19}
\end{equation}
where
\begin{equation}
\ddot{y} (t) + \left\{ \frac{2 \omega (t) \dot{\omega} (t)}{ \omega^2
(t) - 1} - i \left( \omega^2 (t) + 1 \right) \right\} \dot{y} (t) -
\frac{ \left( \omega^2 (t) - 1 \right)^2 }{4} y(t) = 0. \label{3.20}
\end{equation}
Again, the other solutions  are  obtained by substituting Eq. (\ref{3.19})
into Eq. (\ref{3.18})
and integrating.

\section{SQUEEZE OPERATOR AND COHERENT STATES}

\subsection{Time-Independent Harmonic Oscillator}

We review briefly the spectrum-generating algebra of $SU(1,1)$
for the time-independent anharmonic
oscillator of Eq. (\ref{2.1}).
In order to find the exact eigenstates, we rescale the  basis of
Eq. (\ref{2.4}):
\begin{eqnarray}
\hat{K}_0 = \frac{1}{2} \left( \frac{\omega_0 \hat{q}^2 + \hat{p}^2 /
\omega_0}{2} + \frac{c}{\omega_0 \hat{q}^2} \right),  \nonumber \\
\hat{K}_{\pm} = \frac{1}{2} \left( \frac{ \omega_0 \hat{q}^2 - \hat{p}^2 /
\omega_0}{2} - \frac{c}{\omega_0 \hat{q}^2 } \mp i \frac{\hat{p} \hat{q} +
\hat{q} \hat{p} }{2} \right).  \label{4.1}
\end{eqnarray}
The time-independent anharmonic oscillator can now be rewritten simply as
\begin{equation}
\hat{H} = 2 \omega_0 \hat{K}_0. \label{4.2}
\end{equation}
The basis in Eq. (\ref{4.1}) still has the same group structure as that
of Eq. (\ref{2.5}) and forms the spectrum-generating
algebra of $SU(1,1)$ for the
time-independent anharmonic oscillator. It is
Eq. (\ref{4.2}) together with Eq. (\ref{2.5}) that
determines the eigenvalues and group-theoretically
generates the spectrum of eigenstates of the  anharmonic oscillator.

The Casimir operator for the basis of Eq. (\ref{4.1}) is
\begin{equation}
\hat{C} = \frac{1}{2} \hat{K}_0^2 - \frac{1}{4} \left( \hat{K}_+ \hat{K}_-
 + \hat{K}_- \hat{K}_+ \right) . \label{4.3}
\end{equation}
Using the representation in Eq. (\ref{4.1}), the Casimir operator becomes
\begin{equation}
\hat{C} = k_0 (k_0 - 1) \hat{e} = - \left( \frac{3 - 8 c}{16} \right) \hat{e},
\label{4.4}
\end{equation}
where $\hat{e}$ is the identity element,
so we have two unitary irreducible representations
$D^\dagger ( k_0 ) $ that have
positive discrete series \cite{17}  with Bargmann indices \cite{18}
\begin{equation}
k_0 = \frac{1}{2} \left( 1 \pm \sqrt{\frac{1}{4} + 2c} \right), \label{4.5}
\end{equation}
and consist of the eigenstates of $\hat{K}_0 $,
\begin{equation}
\hat{K}_0 \left| n, k_0 \right> = ( n+ k_0) \left| n,k_0 \right>,
\label{4.6}
\end{equation}
where $ n = 0,1,2,\cdots $  The positivity of the Bargmann index
is ensured by the
condition $ c >  -1/8 $ that prevents
the wave functions from falling into the center of
motion \cite{19}. In the pure harmonic oscillator
representation $ (c=0),$ the
representation with the Bargmann index $k_0 = 1/4 $
corresponds to even photon-number
states and $k_0 = 3/4 $ to odd photon-number states
\cite{20}. The operators
$\hat{K}_+$ and $\hat{K}_-$ acting on
$\left| n,k_0 \right>$ increase and decrease
the eigenvalues by one unit; thus, they behave as the raising and the lowering
operators of the number-type eigenstates.
After some algebra, the number-type states are found to be \cite{20}
\begin{equation}
\left| n,k_0 \right> = \left( \frac{\Gamma (2 k_0)}{n! \Gamma (n+2k_0 )}
\right)^{1/2} (\hat{K}_+ )^n \left| 0,k_0 \right>~~. \label{4.7}
\end{equation}

Now, we introduce two definitions of the coherent state.
First, following Perelomov
\cite{21}, the coherent state defined as an eigenstate of $\hat{K}_-$,
\begin{equation}
\hat{K}_- \left| z,k_0 \right> = z \left| z,k_0 \right>, \label{4.8}
\end{equation}
is
\begin{equation}
\left| z,k_0 \right> = \left( 1 - |z|^2 \right)^{k_0} \sum_{n=0}^{\infty}
\left( \frac{\Gamma (2 k_0)}{n! \Gamma (n + 2 k_0 )} \right)^{1/2} z^n
\left|n,k_0 \right>. \label{4.9}
\end{equation}
Second, the coherent state is defined inequivalently as
\begin{equation}
\left| z,k_0 \right> = \hat{S} (\zeta) \left| 0,k_0 \right> , \label{4.10}
\end{equation}
where
\begin{equation}
\zeta = \bigl( \tanh^{-1} (z z^*)^{1/2} \frac{z}{z^*} \bigr)^{1/2},
\label{4.11}
\end{equation}
and
\begin{equation}
\hat{S} (\zeta) = \exp{ \left( \zeta \hat{K}_+ - \zeta^* \hat{K}_- \right)}.
\label{4.12}
\end{equation}
The operator in Eq. (\ref{4.12}) is the squeeze
operator \cite{22}. The coherent
state in Eq. (\ref{4.10}) is  obviously      not
an eigenstate of $\hat{K}_-$. Using the disentangling
technique, one can express the squeeze operator in the form \cite{23}
\begin{equation}
\hat{S}(\zeta) = \exp{ \left( z \hat{K}_+ \right)} \exp{\left( \log( 1 -
|z|)^2 \hat{K}_0 \right)} \exp{ \left(-z^* \hat{K}_- \right)}, \label{4.13}
\end{equation}
and thereby the coherent states as
\begin{equation}
\left|z,k_0 \right> = \left( 1- |z|^2 \right)^{k_0} \sum_{n=0}^{\infty}
\left( \frac{\Gamma (n + 2 k_0 )}{n! \Gamma (2 k_0 )} \right)^{1/2} z^n
\left| n,k_0 \right>. \label{4.14}
\end{equation}
It should be noted that the coherent state in Eq. (\ref{4.8})
defined as an eigenstate of the
lowering operator is not equivalent to that in Eq. (\ref{4.14})
defined by the squeeze
operator. This is in strong contrast with the harmonic
oscillator case for which the
corresponding group is the Heisenberg group
and the two representations are
equivalent \cite{20}. The coherent states of $SU(1,1)$
are discussed with an
emphasis on the harmonic oscillator representation in Ref. \cite{24}.

In coordinate representation, the orthonormal
eigenfunction and energy eigenvalue
are found in terms of the Laguerre polynomials \cite{25} as
\begin{eqnarray}
\Phi_n (q) &=& \left( \frac{\omega_0^{1/2} n!}{\Gamma (n + 2 k_0 )^3}
 \right)^{1/2}
\exp{(- \omega_0 q^2 / 2)} (\omega_0 q^2 )^{k_0 - 1/4}
L^{(2k_0 -1)}_n ({\omega}_0 q^2 ),
\nonumber \\
E_n &=& 2 \omega_0 (n+k_0 ) . \label{4.15}
\end{eqnarray}
Equation ( \ref{4.15}) agrees with the group-theoretical result of
Eq. (\ref{4.6}), as expected.

\subsection{Time-Dependent Anharmonic Oscillator}

The exact quantum states of the time-dependent harmonic oscillator of
Eq. (\ref{3.6})
are determined from the number states of its generalized invariant up to some
time-dependent phase factors. Similarly,
the exact quantum states of the time-dependent anharmonic
oscillator  can be found
from the number-type eigenstates of the generalized invariant in
Eq. (\ref{3.4}) up to
some time-dependent phase factors. This is one of the great advantages of the
generalized invariant method.

In order to find the eigenstates, first let us
introduce a canonical transformation
\begin{equation}
\hat{q}_c = \hat{q} ,~~ \hat{p}_c = \hat{p} + \frac{g_0 (t) \hat{q} }{g_- (t)
},
\label{4.16}
\end{equation}
and rewrite the generalized invariant as
\begin{equation}
\hat{I} (t) = \frac{g_- (t) \hat{p}^2_c }{2} + \frac{\omega^2_0 \hat{q}^2_c}{2
g_- (t) } + \frac{c g_- (t)}{\hat{q}^2_c}             \label{4.17}
\end{equation}
where
\begin{equation}
\omega^2_0 = g_+ (t) g_- (t) - g_0^2 (t) \label{4.18}
\end{equation}
is a constant of motion. The constancy is a consequence of the
invariance of Eq. (\ref{3.3}) \cite{3}.

In analogy with Eq. (\ref{4.1}) we now introduce a time-dependent basis
of $SU(1,1)$ as
\begin{eqnarray}
\hat{K}_{c,0} (t) &=& \frac{1}{2} \left( \frac{\omega_0
\hat{q}^2_c / g_- (t) + g_-(t)
\hat{p}_c^2 / \omega_0 }{2} + \frac{c g_- (t) }{\omega_0 \hat{q}^2_c }
\right), \nonumber \\
\hat{K}_{c,\pm} (t) &=& \frac{1}{2} \left( \frac{ \omega_0 \hat{q}^2_c / g_-
(t)
- g_- (t) \hat{p}^2_c / \omega_0 }{2} - \frac{c g_- (t) }{\omega_0 \hat{q}^2_c}
\mp i \frac{\hat{p}_c \hat{q}_c + \hat{q}_c \hat{p}_c }{2} \right) .
\label{4.19}
\end{eqnarray}
Of course, it holds that $\hat{K}^*_{c,+} = \hat{K}_{c,-}$. The
time-dependent anharmonic oscillator can still be rewritten as
\begin{equation}
\hat{I} (t) = 2 \omega_0 \hat{K}_{c,0} (t) . \label{4.20}
\end{equation}
We obtain the Casimir operator just by replacing the basis of Eq. (\ref{4.1})
with the basis of Eq. (\ref{4.19}) in Eq. (\ref{4.3}).
The Bargmann index, however, does not change and is
given by the same value as in
Eq. (\ref{4.5}). The only modification is the number-type eigenstate
\begin{equation}
\hat{K}_{c,0}(t)\left| n,k_0,t \right> = (n+k_0) |n,k_0,t>. \label{4.21}
\end{equation}
One of the properties of the generalized invariant is that the eigenvalues
are always time-independent, whereas the eigenstates are time-dependent.
By applying the raising operators we obtain the number-type eigenstate
\begin{equation}
|n,k_0,t> = \left(\frac{\Gamma (2k_0)}{n! \Gamma(n+2k_0)}\right)^{1/2}
(\hat{K}_{c,+})^n |0,k_0,t> \label{4.22}
\end{equation}
and the coherent state
\begin{equation}
\left|z,k_0,t \right> = (1-|z|^2)^{k_0} \sum_{n=0}^{\infty}
\left( \frac{\Gamma(2k_0)}{n!\Gamma(n+2k_0)}\right)^{1/2} \left|n,k_0,t\right>.
\label{4.23}
\end{equation}

After a bit of algebra gymnastics, we find that
\begin{equation}
\frac{\partial}{\partial t}\hat{K}_{c,t}(t) = - \frac{1}{2g_-(t)}
\frac{dg_-(t)}{dt} \hat{K}_0 (t) - i \frac{g_- (t)}{2 \omega_0 } \frac{d}{dt}
\left( \frac{g_0 (t) }{ g_- (t)} \right) \left( \hat{K}_{c,+} (t) -
\hat{K}_{c,-} (t) \right)    \label{4.24}
\end{equation}
and
\begin{equation}
\left< n,k_0,t \right| \frac{\partial}{ \partial t} \left| n,k_0,t \right>
= - i \frac{g_- (t)}{2 \omega_0 } \frac{d}{dt} \left( \frac{ g_0 (t)}{g_- (t)}
\right) (n+k_0 ). \label{4.25}
\end{equation}
The expectation value of the Hamiltonian of the anharmonic oscillator in
Eq.~(\ref{3.1}) is also given by
\begin{equation}
\left< n,k_0,t \right| \hat{H} (t) \left| n,k_0,t \right> = \frac{ \omega_0^2 +
g_0^2 (t) + \omega^2 (t) g_-^2 (t) }{\omega_0 g_- (t) } (n + k_0).
\label{4.26}
\end{equation}
Equations (\ref{4.25}) and (\ref{4.26}) for
the anharmonic oscillator are the same as
Eq. (36) in Ref.[9.(b)] and Eqs. (\ref{4.6}) and
(\ref{4.8}) in Ref.[9(c)] for the
harmonic oscillator of Eq. (\ref{3.6}),
except for a modification by $(n + k_0)$.
Equation (\ref{4.24}), however, differs from Eq. (4.5)
in Ref.[9(c)] because in our
case the relevant algebra that was used to construct
the number-type eigenstate
was that of $SU(1,1)$ compared to the Heisenberg
group of the creation and
the annihilation operators for a harmonic oscillator.
It is a well-known result of
the generalized invariant that the exact quantum states of the anharmonic
oscillator are
\begin{equation}
\left| \psi, n, k_0 , t \right> = \exp{\bigl(-i \int \left( h(t) - i \epsilon
(t) \right) (n + k_0 )\bigr) } \left| n, k_0,t \right> \label{4.27}
\end{equation}
where
\begin{equation}
h(t) = \frac{ \omega_0^2 + g_0^2 + \omega^2 (t) g_-^2 (t)}{\omega_0 g_- (t)
},~~ \epsilon (t)  = - i \frac{g_- (t) }{ 2 \omega_0 } \frac{d}{dt} \left(
\frac{g_0 (t) }{g_- (t) } \right). \label{4.28}
\end{equation}

By replacing $\omega_0 q^2 $ with $ \omega_0 q^2 / g_- (t) $ in
Eq. (\ref{4.15}), one obtains straightforwardly
the orthonormal eigenfunctions
\begin{equation}
\Phi_n (q,t) = \left( \frac{\omega^{1/2} n! }{ \Gamma (n + 2 k_0 )^3}
\right)^{1/2} \exp{\left(- \omega_0 q^2 / 2 g_- (t) \right) }
\left( \frac{\omega_0 q^2}{g_-(t)} \right)^{k_0 - 1/4} L_n^{(2 k_0 -1)}
\left( \frac{ \omega_0 q^2}{g_- (t)} \right).     \label{4.29}
\end{equation}
Furthermore, with the generalized canonical transformation
\begin{equation}
\hat{\tilde{q}} = \frac{\hat{q}}{\rho},~~ \hat{\tilde{p}} = \rho \hat{p} -
\dot{\rho} \hat{q}, \label{4.30}
\end{equation}
one can transform the generalized invariant of Eq. (\ref{3.4})
into a time-independent
one:
\begin{equation}
\hat{I} (t) = \frac{1}{2} \left( \hat{\tilde{p}}^2 + \hat{ \tilde{q}}^2 \right)
+ \frac{c}{ \hat{\tilde{q}}^2}. \label{4.31}
\end{equation}

\section{EVOLUTION OPERATOR METHOD}

In Sec. III, we have seen that the time-dependent anharmonic oscillator
of Eq. (\ref{3.1}) preserves the
Lie algebra of $SU(1,1)$ with the time-dependent basis
of Eq. (\ref{4.19}) during its
evolution, and that enables us to find the exact quantum states
group-theoretically. In this section, we shall investigate systematically the
evolution of the basis.

We assume that
\begin{eqnarray}
\hat{H} &=& \frac{\hat{p}^2}{2} + \frac{\omega^2_0 \hat{q}^2 }{2} + \frac{c}{
\hat{q}^2 } ,~~ t \leq t_0 , \nonumber \\
\hat{H} (t) &=& \frac{ \hat{p}^2 }{2} + \frac{ \omega^2 (t) \hat{q}^2}{2}
+ \frac{c}{ \hat{q}^2} ,~~ t \geq t_0 ,   \label{5.1}
\end{eqnarray}
for an initial time $t_0$. Then, it is easy to see that
\begin{equation}
\hat{K}_{c,0} (t) = u_0 (t) \hat{K}_0 + u_+ (t) \hat{K}_+ + u_- (t) \hat{K}_-
\label{5.2}
\end{equation}
with
\begin{eqnarray}
u_0 (t) &=& \frac{1}{2} \left( g_- (t) + \frac{1}{g_- (t)} +
\frac{ g_0^2 (t)}{\omega_0^2 g_-(t)} \right), \nonumber \\
u_{\pm} (t) &=& \frac{1}{4} \left(  -g_- (t) + \frac{1}{g_- (t)} \pm 2 i
\frac{g_0 (t)}{\omega_0} + \frac{g_0^2 (t) }{ \omega^2_0 g_- (t)}
\right), \label{5.3}
\end{eqnarray}
and
\begin{equation}
\hat{K}_{c,+} (t) = \nu_0 (t) \hat{K}_0 + \nu_+ (t) \hat{K}_+ + \nu_- (t)
\hat{K}_- \label{5.4}
\end{equation}
where
\begin{eqnarray}
\nu_0 (t) &=& \frac{1}{2} \left( -g_- (t) + \frac{1}{g_-(t)} -i \frac{g_0(t)}
{\omega_0 g_-(t)} -
\frac{g_0^2 (t)}{\omega_0^2 g_-(t)} \right), \nonumber \\
\nu_{\pm} (t) &=& \frac{1}{2} \left( g_- (t) \pm \frac{1}{g_- (t)} \mp i
\frac{g_0 (t)}{2\omega_0 g_-(t)}
\mp i \frac{g_0(t)}{\omega_0}
 \mp \frac{g_0^2 (t)}{2 \omega_0^2 g_- (t)}
\right),    \label{5.5}
\end{eqnarray}
and $ \hat{K}_{c,-} (t) = \hat{K}^*_{c,+} (t)$. Before the initial time,
the generalized invariant in Eq. (\ref{3.4}) is simply the Hamiltonian
of Eq. (\ref{2.1})
itself, $g_{\pm} (t_0) = 1$, and $g_0 (t_0) = 0$; so $ u_0 (t_0) = \nu_+
(t_0) = 1 $, and all the other terms vanish, as expected.

Since $u_(t) = u^*_+ (t)$ , one may write $\hat{K}_{c,0} (t) $ as
\begin{equation}
\hat{K}_{c,0} (t) = \hat{S}^\dagger (\xi) \hat{K}_0 \hat{S} (\xi)
\label{5.6}
\end{equation}
where the squeeze operator
\begin{equation}
\hat{S} (\xi) = \exp{(\xi \hat{K}_\dagger - \xi^* \hat{K}_-)} \label{5.7}
\end{equation}
has the squeeze parameter
\begin{equation}
\left(\xi \xi^* \right)^{1/2} = \frac{1}{2} \tanh^{-1} \bigl(\frac{2(u_+
u^*_+)^{1/2}}{u_0} \bigr) ,~~ \frac{\xi^*}{\xi} = \frac{u_+^*}{u_+}.
\label{5.8}
\end{equation}
It follows then that the vacuum state at an arbitrary later time is the
coherent state of the initial vacuum state
\begin{equation}
\left| 0, k_0,t \right> = \hat{S}^\dagger (\xi) \left| 0,k_0,t_0
\right> \label{5.9}
\end{equation}
and that the number-type eigenstate is again the coherent state of the same
initial number-type eigenstate
\begin{equation}
\left| n,k_0 ,t \right> = \hat{S}^\dagger (\xi) \left| n,k_0,t_0
\right>.
\end{equation}
In the case of the time-dependent harmonic oscillator,
it was pointed out in
Refs. \cite{8} and \cite{26} that the exact quantum states
are the squeeze states of
the initially prepared quantum states. This also holds for
the time-dependent anharmonic oscillator.

In order to see how an initially prepared quantum state
evolves into a quantum state an arbitrary
later time, one can also use the evolution operator obtained in Sec. III.
In spite of the facts that both of the evolution operators,
Eqs. (\ref{3.13}) and
(\ref{3.14}), are the same from the uniqueness of the evolution operator
for the evolution equation ( Eq. (\ref{3.12})) and
that it is easier to solve
Eq. (\ref{3.15}) than Eq. (\ref{3.18}), the evolution operator of
Eq. (\ref{3.14}) is
more useful because the basis of the spectrum-generating algebra of the
anharmonic oscillator is Eq. (\ref{4.1}). The initial state prepared as one
of the number-type eigenstate in Eq. (\ref{4.6}),
\begin{equation}
\hat{H} \left| n,k_0,t_0 \right> = 2 \omega_0 (n+k_0 ) \left| n,k_0,t_0
\right>, \label{5.11}
\end{equation}
evolves at an arbitrary later time according to
\begin{equation}
\left| n,k_0 , t \right> = \hat{U} (t,t_0 ) \left|n,k_0,t_0 \right>
\label{5.12}
\end{equation}
into the same number-type eigenstate
with the same energy eigenvalue $2 \omega_0 (n+k_0 )$ of the generalized
invariant $\hat{I} (t)$. From the evolution of the Hamiltonian in the
Heisenberg picture, it follows that the basis transforms unitarily as
\begin{equation}
\hat{K}_{c,0} (t) = \hat{U} (t,t_0) \hat{K}_0 \hat{U}^\dagger (t,t_0).
\label{5.13}
\end{equation}
One can see that the squeeze operator is the Hermitian conjugate
of the evolution operator:
\begin{equation}
\hat{S} (\xi) = \hat{U}^\dagger (t,t_0). \label{5.14}
\end{equation}
The generalized invariant method gives exactly the same result as the
evolution operator method.

For example, we consider a class of time-dependent harmonic oscillators
\begin{equation}
\hat{H}_0 (t) = \frac{\hat{p}^2}{2} + \frac{\omega_0^2 t^\alpha
\hat{q}^2}{2},  \label{5.15}
\end{equation}
which have the one-parameter-dependent generalized invariant [9(c)]
\begin{eqnarray}
g_- (t)  &=& \frac{\pi}{2} \left( \frac{2 \omega_0}{2 +\alpha}
\right)^{\alpha \nu} c_1 z^{2 \nu} \left[ J^2_{\nu} (z) + N^2_{\nu} (z)
\right], \nonumber \\
g_0 (t) &=& - \frac{\pi \omega_0}{2} c_1 \left[ \left(\nu J_\nu (z) + z J'_\nu
(z) \right)J_\nu (z) + \left( \nu N_\nu (z) + z N'_\nu (z) \right) N_\nu
(z) \right], \nonumber \\
g_+ (t) &=& \frac{\pi \omega_0^{2 \nu + 1}}{2} \left( \frac{2}{2 + \alpha}
\right)^{- \alpha \nu} c_1 z^{-2 \nu} \left[ \left( \nu J_\nu (z) + z
J'_\nu (z) \right)^2 + \left( \nu N_\nu (z) + z N'_\nu (z) \right)^2
\right],  \label{5.16}
\end{eqnarray}
where $ \nu = 1/(2+\alpha), z = 2 \omega_0 t^{(2+\alpha)/2}
/(2+\alpha)$, and $c_1$ is a constant. Then, we are able to find the
generalized invariant for the time-dependent anharmonic oscillator,
\begin{equation}
H(t) = \frac{p^2}{2} + \frac{ \omega_0^2 t^\alpha q^2}{2} +
\frac{c}{q^2},      \label{5.17}
\end{equation}
by substituting Eq. (\ref{5.16}) into Eq. (\ref{3.4}) with the basis of
Eq. (\ref{2.3})
and by rewriting in the basis of Eq. (\ref{2.4}).

\section{DISCUSSION}

The  quantum harmonic oscillator with an inverse squared singular term
is a well-known problem that was exactly solved. In this problem, the singular
term is related to the repulsive potential of the angular momentum in the
radial motion of a three-dimensional isotropic harmonic oscillator, up to
some power of the radius \cite{14,19}. It is also related to
the relative motion of
the two-body Calogero model \cite{12}, a prototype of exactly solvable
models. In spite of its familiarity in physics, the majority of
approaches so far has been analytic rather than group theoretical. Even
the group theoretical approach to this problem has not gone beyond the
spectrum-generating algebra of $SU(1,1)$ and the spectrum of energy
eigenvalues \cite{14}, in contrast with the harmonic oscillator problem
which has been intensively studied group-theoretically based on the
Heisenberg group and $SU(1,1)$.

In this paper, we considered the time-dependent harmonic oscillator
with the inverse squared singular term. Even though the exact
eigenfunctions of the time-dependent anharmonic oscillator were
analytically found in Ref.\cite{11}, we uncovered the underlying
groups $SU(2)$ and $SU(1,1)$, found the generalized invariant, and
constructed the number-type eigenstates and coherent states
group-theoretically, which are the main results of this paper. Based on
the Lie algebra of $SU(2)$, it was observed that the generalized
invariant in Eq. (\ref{3.4}) for the time-dependent anharmonic oscillator
in Eq. (\ref{3.1}) had the same invariant equation (Eq. (\ref{3.6}))
ans was determined by the classical integrals of motion
for the time-dependentharmonic oscillator of Eq.\ref{3.6},as did the
evolution operator in Eq. (\ref{3.13}) \cite{8}.
We showed that the generalized invariant preserved the same spectrum-
generating Lie algebra of $SU(1,1)$ by introducing the time-dependent
basis in Eq. (\ref{4.19}) and had the time-dependent number-type eigenstates
of Eq. (\ref{4.22}) and the coherent states of Eq. (\ref{4.23}).
The exact quantum states
were given by Eq. (\ref{4.27}), whose time-dependent phase factors had the
same form (Eq. (\ref{4.28})), except for the Bargmann index, as those for the
harmonic oscillator. It was shown that the squeeze operator, (Eq. (\ref{5.7}))
transformed unitarily the time-dependent basis according to Eq.
(\ref{5.6}). Therefore, the number-type eigenstate transformed as in Eq.
(\ref{5.12}) by the squeeze operator, and in particular the vacuum at an
arbitrary later time was a coherent state of the initial vacuum. The
eigenstate evolved as a squeeze state of the initial number-type state.
Finally, it was shown that the squeeze operator was the Hermitian
conjugate of the evolution operator. It was manifest that the task to
find the generalized invariant was relatively easier than that to
evaluate the evolution operator directly. As an example, the harmonic
oscillator of Eq. (\ref{5.17}) was worked out explicitly.

Further application of the
generalized invariant method to time-dependent extended harmonic
oscillators with perturbation terms beyond the linear force term and
the inverse squared singular term
may be tremendously difficult, but promising.

\acknowledgments
This work was supported in part by a Non Directed
Research Fund, Korea Research Foundation, 1994 and by the Basic Research
Institute Program of the Korea Ministry of Education under Contract No.
94-2427.

\end{document}